# A SHAKY ROAD TO SUBNANOMETER BEAMS. NLC GROUND MOTION, VIBRATION AND STABILIZATION STUDIES[1]

Andrei Seryi, SLAC, Stanford, USA

*Abstract*

Ground motion and vibration can be a limiting factor in the performance of future linear colliders. Investigations of ground motion have been carried out around the world for several decades. In this review, results of recent investigations of ground motion as well as ongoing developments of stabilization methods are presented.

## 1 GROUND MOTION STUDIES

Ground motion can conveniently be divided into 'fast' and 'slow' when studying its effect on a linear collider. Fast motion (f > a few Hz) cannot be adequately corrected by a pulse-to-pulse feedback operating at the repetition rate of the collider and therefore primarily causes beam offsets at the IP. Slow motion can be compensated by feedback and thus results only in beam emittance growth. Another consideration is that the mechanism that produces relative displacements is different (discussed below) for slow and fast motion with a boundary occurring somewhere in a milli-Hz range.

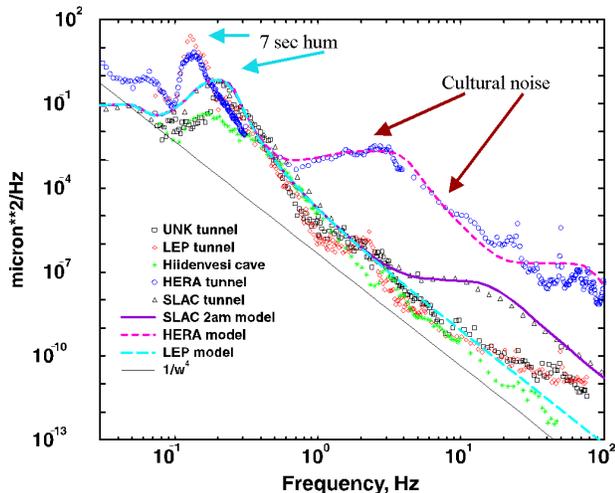

Figure 1: Power spectrum of absolute ground motion measured in different sites [1,2,3,4,5].

Fast motion is typically represented by a power spectrum as shown in Figure 1. The power spectrum of absolute displacement generally follows a $1/\omega^4$ behavior on top of which are added the particular signatures from the contribution of different sources. For example, cultural noise (human produced) typically manifests itself above 1 Hz. Motion from ocean waves causes a peak around 0.14 Hz.

The integrated spectrum over a particular frequency band gives the corresponding RMS motion. It is easy to see that the natural ground motion is quite small, as low as a fraction of a nanometer, for frequencies higher than several Hz (see Figure 2).

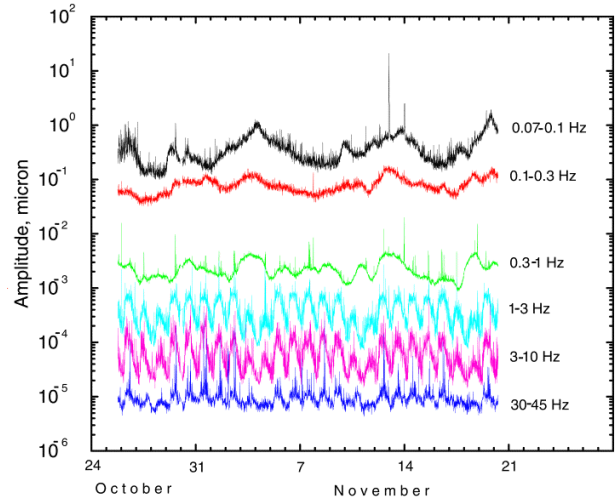

Figure 2: Ground motion measured in Hiidenvesi cave showing the amplitude for different frequency bands [5].

The motion in the low frequency bands in Figure 2 is much larger, the order of a micrometer, however it is important to emphasize that the amplitude shown is absolute, i.e. it is the motion of a single point with respect to an inertial reference frame. What is important for a linear collider is the relative motion of two quadrupoles separated by distances less than the betatron wavelength.

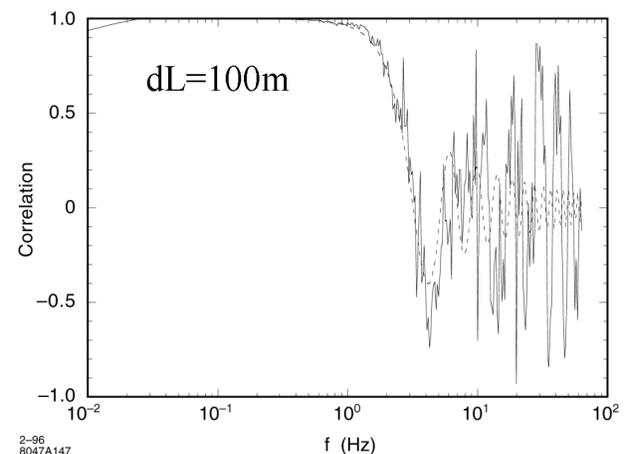

Figure 3: Example of correlation measured in the SLC tunnel for 100 m separation between probes [4].

[1] Work supported by the U.S. Department of Energy, Contract DE-AC03-76SF00515.

Correlation measurements have shown [1,2,4] that natural ground motion consists mostly of elastic waves with a wavelength given by the phase velocity in the media. The slowest band of 'fast' motion has quite a long wavelength and is therefore highly correlated (see Figure 3) and does not cause harmful effects on a linear collider.

In the NLC, the tolerance on jitter in the position of elements varies significantly. The tolerance for the Final Doublet (FD) is a fraction of the beam size at the IP, around 1 nm. Some of the quadrupoles in the beam delivery system have very tight tolerances as well, on the order of 5-10 nm. The jitter tolerances for the main linac quadrupoles are also of the order of 10 nm, since there are many of them. The relevant frequency range for these tolerances is determined by the beam-based feedback which would effectively compensate for motion slower than about 1/20 of the repetition rate; for the JLC/NLC design, the frequency band of concern is f > 6 Hz. More detailed evaluations of tolerances are given in the next section.

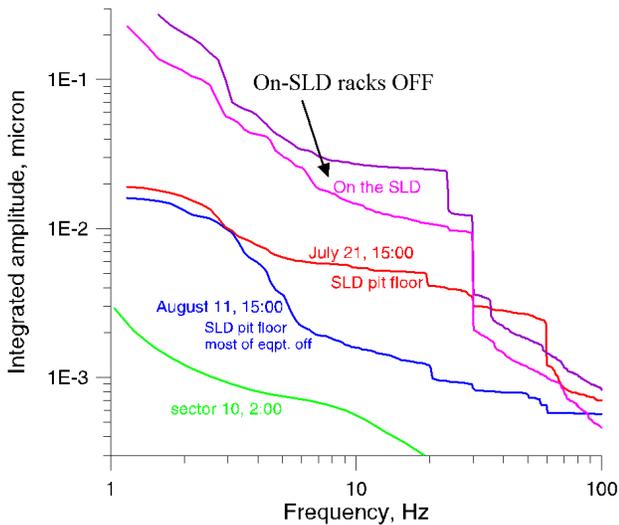

Figure 4: Integrated spectrum of absolute motion in SLC sector 10 compared with the motion near the SLD detector for a variety of situations. The top curves show motion on the SLD with and without the electronic racks on. The lower curves show the noise on the pit floor under nominal conditions and when most of the noise sources in the building were turned off (SLD door opened, most of on-SLD racks turned off) [6].

From these considerations, one can conclude that natural fast ground motion does not represent a limitation for a linear collider (for present parameters) and the real concern is cultural noise produced in the vicinity, whether external or internal to the tunnel or vibrations produced on the accelerator girder itself. An example of cultural noise is shown in Figure 4 where the noise measured on the floor of the SLC collider hall and on the SLD detector is shown in comparison with a quiet location at SLAC. Most of the noise is produced by very local sources – building ventilation, nearby compressors, etc. – and by various equipment mounted on the SLD itself. It is clear that the NLC detector must be designed carefully with a goal of minimizing vibration. Moreover, all of the conventional facilities support equipment for the accelerator will have to satisfy vibration criteria.

The fast motion and vibration discussed above is not the only issue for a linear collider. Ground motion below 0.01 Hz or so, in spite of being very slow, can have a rather short wavelength, causing misalignments of the collider and producing emittance growth. This motion is not wave-like and can be inelastic. There are two types of motion – one is diffusive and another is systematic. The model for the diffusive motion parameterizes the RMS relative misalignment as an ATL law [2,8]: $\Delta X^2 = ATL$ where T is the time since perfect alignment and L is the distance between points.

The parameter A varies by a few orders of magnitude when measured in different places and is clearly site and geology dependent. For example, measurements at the DESY site gave a value of $A = 10^{-5}$ μm$^2$/m/s [9] while in a tunnel built in rock in Japan, $A = 2*10^{-9}$ μm$^2$/m/s. The latter measurements also showed an increase in the value of A when blasting was used for the tunnel construction [10]. For diffusive motion with a value of $A = 5*10^{-7}$ μm$^2$/m/s, similar to observations at SLAC in the SLC and FFTB tunnels [11,7], the linear collider would require continuous beam based alignment on a time scale of every 30 minutes in order to prevent emittance growth.

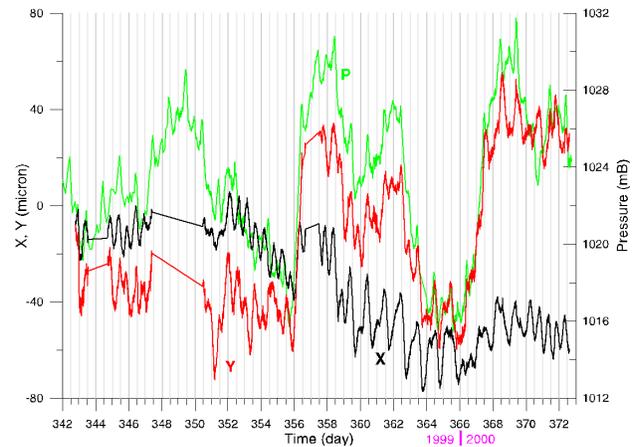

Figure 5: Horizontal and vertical motion of the center of the 3 km SLAC tunnel with respect to its ends plotted along with external atmospheric pressure [7].

Recent measurements at SLAC have shown that the value of A can depend on changes in atmospheric pressure acting on the ground [7]. Figure 5 shows data taken in the 3 km SLAC tunnel over a period of one month. There is a clear tidal component seen in the transverse motion of the tunnel and there is a clear correlation of the motion with external atmospheric pressure. This correlation was observed over a wide frequency band (3E-6 – 1E-3 Hz) where the power spectrum of the tunnel motion behaved as $1/\omega^2$ (the same as the ATL law) and thus the value of A could be derived.

The spectrum of atmospheric pressure was found to behave in a similar way (also as $1/\omega^2$) and its amplitude $A_p$ was surprisingly correlated with A as shown in Fig.6.

The observation that the deformation of the linac tunnel is correlated with atmospheric pressure variation [7] can be explained by assuming that the pressure acts on the ground whose properties vary along the linac. This variation can be due to changes in the Young's modulus E, changes in the topology of the surface, or changes in the characteristic depth of the softer surface layers. It is worth noting that in this case the atmosphere-driven contribution to A scales as $1/E^2$ and therefore depends strongly on geology which may be partly responsible for the large variation of A observed in different places.

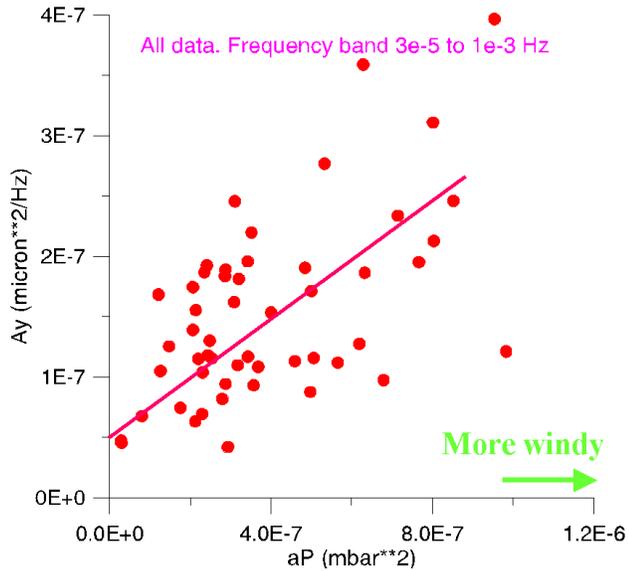

Figure 6: Parameter A of the ATL law defined from vertical motion of SLC tunnel plotted against the amplitude $A_p$ of the atmospheric pressure spectrum [7].

Very slow ground motion can also be systematic in time over periods of months to years. Such motion has been observed at SLAC, CERN and other places [12]. In some cases such motion can also be described by a simple rule for its RMS relative misalignment [13]: $\Delta X^2 = A_S T^2 L$ with another site-specific coefficient $A_S$. At SLAC it appears that the systematic motion dominates on a time scale greater than $10^5$ seconds. Studies to investigate this slow ground motion in more detail are currently in progress at many different laboratories.

## 2 MODELING

In order to accurately characterize the influence of ground motion on a linear collider, a comprehensive mathematical model is required. Such a model would include understanding of the temporal and spatial properties of the motion and of the driving mechanisms. In most cases an adequate representation for such a model consists of the 2-D power spectrum $P(\omega,k)$ based on measured spectra of absolute motion and correlation.

The ground motion model for the SLAC site [13] uses measurements of fast motion taken at night in sector 10 of the SLAC linac [4]. Figure 7 shows the spectrum of absolute and relative motion for this model. To evaluate different levels of cultural noise, we augment this model to include two other cases with significantly higher and lower noise contributions. The measured spectra and the approximations used in the models are shown in Figure 1.

The "HERA model" is based on measurements in DESY [3] and corresponds to a very noisy shallow tunnel located in a highly populated area where no precautions were taken to reduce the contribution of various noise sources in the vicinity or in the tunnel. The "LEP model" corresponds to a deep tunnel where the noise level is very close to the natural seismic level, without additional cultural sources outside or inside of the tunnel. The "SLAC model" represents a shallow tunnel located in a moderately populated area.

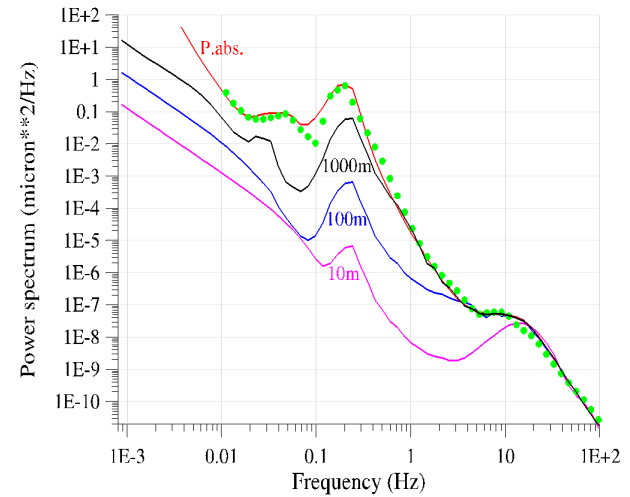

Figure 7: Measured (points) and modeling spectrum $p(\omega)$ of absolute motion for the 2 AM SLAC ground motion model [4,13]. The lower curves show the spectra of relative motion $p(\omega,L)/2$ for different separation L.

As an example of how such models can be used, a comparison of the performance of the NLC Final Focus for different ground motion models is shown in Figure 8. One can see that a site located in a highly populated area without proper vibration sensitive engineering would present significant difficulties for a linear collider with the parameters considered. Stabilization of more components than just the final doublet would be necessary. A site with noise similar to the "SLAC model" would certainly be suitable, while the "LEP model" would be suitable even for much more ambitious beam parameters. These results should not be considered as an attempt to evaluate any particular site, or even the models, because for a fully consistent assessment, various in-tunnel noise sources as well as vibration compensation methods must be considered as an integrated system.

The SLAC ground motion model includes all of the features that are presently known, including wave motion,

diffusive motion and systematic motion. Figure 9 shows how these types of motion affect the behavior of RMS relative misalignment over time. One can see that this curve can be divided into three regions: wave dominated, ATL-dominated and systematic motion dominated.

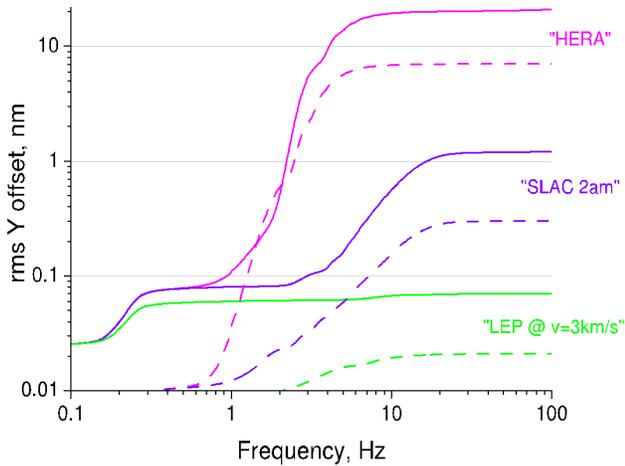

Figure 8: Integrated spectral contribution to the RMS equilibrium IP beam offset for the NLC Final Focus with final doublets supports at ±8 m for different models of ground motion. Dashed curves correspond to the case when the relative motion of the FDs is eliminated [14].

## 3 STABILIZATION METHODS

For the NLC, the ground motion and vibration issues have been systematically addressed by proper design, careful consideration of sites and geology, and by development of appropriate stabilization methods.

An important issue for the NLC design is the location. An ideal site should have little external cultural noise, now or in the future. Solid rock is the preferred surrounding media since fast motion will be better correlated and slow motion will be reduced. In real life the proximity to an existing major laboratory would also be a great advantage. This proximity would not necessarily mean compromising the other requirements since a deep tunnel version of the NLC could have good geology and low noise and still be located near an existing lab. This is true for all of the deep tunnel NLC locations that have been considered in both Illinois and California.

Of course a good site alone is not sufficient, the noise generated by the linear collider equipment itself and by conventional facilities equipment must be appropriately controlled and minimized (by design and by further passive or active damping). This would allow the tolerances to be met without any additional active stabilization or correction techniques for all of the focusing and accelerating elements of the linear collider with the only exception being the final doublet.

Though the NLC detector must be designed to minimize vibrations, it is unlikely that the tolerances for the final quadrupoles mounted on the detector could be met without additional active measures. Several methods are being developed to provide the necessary relative stability for the final doublet, among them position stabilization via feedback and correction of the magnetic center position with dipole coils via feedforward. Both methods would rely on either inertial measurements of the motion by seismometers or on optical interferometric measurements of their position with respect to each other or to stable ground under the detector ("Optical anchor").

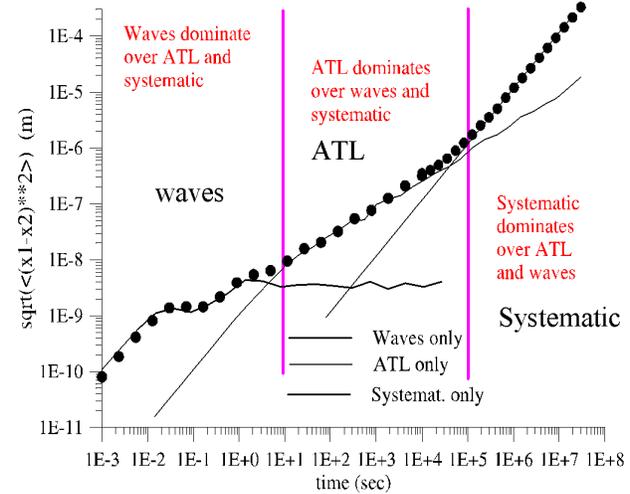

Figure 9: RMS relative motion versus time for two points separated by 30 m for the 2 AM SLAC site ground motion model [13].

The "Optical anchor" developed first at SLAC [15] and now being expanded at UBC [21] has demonstrated a resolution of about 0.2 nm [15] which would be more than sufficient for measuring the position of the FD.

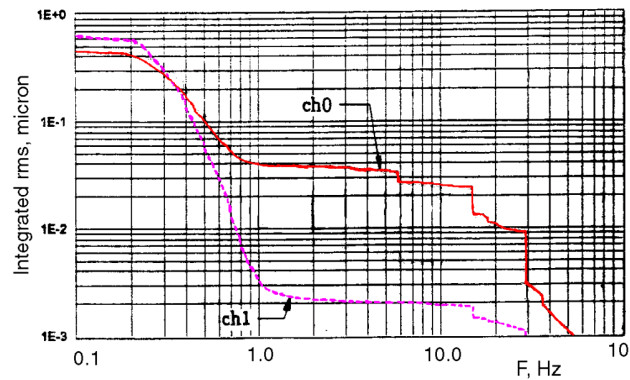

Figure 10: Integrated spectrum measured on the floor (ch0) and on top of a 1500 kg PEP-II quadrupole (ch1) stabilized with the use of three STACIS vibration insulation stands [18,20].

An attempt to stabilize the position of a quadrupole was made at DESY as part of the S-Band linear collider project [16]. In this case a single seismometer and a single piezo-mover were used to stabilize the effective position of the quadrupole center. A reduction of RMS motion by a factor of 3 was achieved (from 100 nm to about 30 nm for frequencies higher than a few Hz).

Another attempt at inertial stabilization was performed at SLAC using three commercial STACIS insulation stands now produced by TMC [17] to stabilize the position of a 1500 kg PEP-II quadrupole [18,20]. In this case the floor motion was reduced by about a factor of 20 (from 40 nm to 2 nm for f > 2 Hz). However, additional slow noise of the order of 200 nm was introduced (see Figure 10) and the performance in the horizontal plane was not as satisfactory.

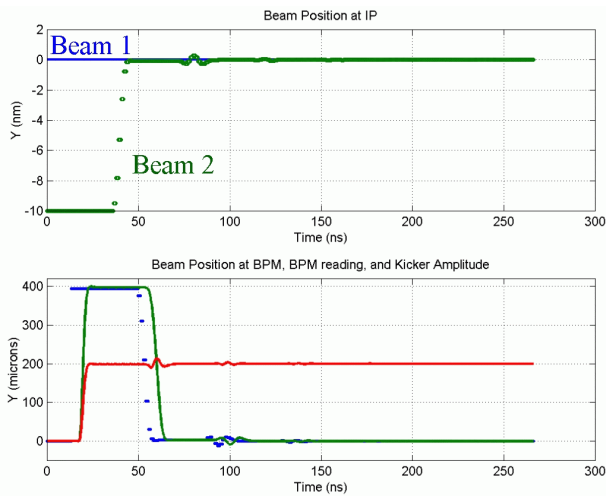

Figure 11: Simulated performance of the fast intra-bunch feedback showing the capture transient for an initial beam-beam offset of about three sigma. The upper plot shows the position of the beams at the IP. The lower plot shows the position of the outgoing beam at the BPM (blue), the analog response of the BPM processor (green), and the kicker drive signal, in arbitrary units (red). Essentially full luminosity is restored in 42 ns [24].

These first examples of inertial stabilization (or inertial sensing for feedforward) do not satisfy the requirements for NLC. In a real collider, the system must detect and appropriately minimize motion of two extended and separated FDs without excessively disturbing the correlation with the rest of machine. In addition, the system must work in an external magnetic field and it must be compact and reliable. R&D to address these issues will continue at SLAC and in other laboratories [21,22,23].

In addition to the beam independent methods of IR stabilization, beam based methods are also being developed, in particular a fast correction within the bunch train. The intra-train feedback will use a position monitor (BPM) near the IP to detect the offset of the first bunches of the train. The signal is the beam-beam deflection due to the relative offset of the beams and fast kickers are then used to correct the rest of the bunch train. Recent preliminary evaluation indicates that such system is technically feasible with available components and could provide efficient capture of beams with several sigma offset [24]. This would significantly reduce the requirements on incoming beam jitter and on stabilization of the FD. Further R&D will concentrate on a technical demonstration of this system.

## CONCLUSION

Ground motion and vibration are important issues for any future linear collider and they have been extensively studied and modelled at different laboratories around the world. In this paper, we have outlined our present understanding of the ground motion and vibration problem and described the models used to evaluate the effects. We have also discussed the ongoing efforts on stabilization techniques that are being pursued by the NLC group.

The review presented is certainly just a brief snapshot of this very actively developing field. To get a broader picture of the issues, including discussion of similar problems in circular accelerators as well as in non-accelerator experiments, the reader is referred to the proceedings of the recent ICFA Workshop on Ground Motion held at SLAC in November 2000 [19].

This paper summarizes results obtained by many people from laboratories throughout the world. The author would like to acknowledge all of their contributions.